\documentclass[aps, prb,twocolumn,showpacs,preprintnumbers,amsmath,amssymb]{revtex4}

\usepackage{graphicx}
\usepackage{dcolumn}
\usepackage{bm}

\begin{document}

\preprint{PRB/ZAX-0410}

\title{Nernst Effect and Superconducting Fluctuations
\\in Zn-doped YBa$_{2}$Cu$_{3}$O$_{7-\delta}$}

\author{Z. A. Xu }

 \email{zhuan@css.zju.edu.cn}
\author{J. Q. Shen}
\author{S. R. Zhao}
\affiliation{%
Department of Physics, Zhejiang University, Hangzhou 310027, P. R.
China
}%

\author{Y. J. Zhang, C. K. Ong}
\affiliation{ Department of Physics, National University of
Singapore, Low Kent Ridge Road, Singapore 119260
}%

\date{\today}

\begin{abstract}
We report the measurements of in-plane resistivity, Hall effect,
and Nernst effect in Zn doped YBa$_{2}$Cu$_{3}$O$_{7-\delta}$
epitaxial thin films grown by pulsed laser deposition technique.
The pseudogap temperature, $T^*$, determined from the temperature
dependence of resistivity, does not change significantly with Zn
doping. Meanwhile the onset temperature ($T^{\nu}$) of anomalous
Nernst signal above $T_{c0}$, which is interpreted as evidence for
vortex-like excitations, decreases sharply as the superconducting
transition temperature $T_{c0}$ does. A significant decrease in
the maximum of vortex Nernst signal in mixed state is also
observed, which is consistent with the scenario that Zn impurities
cause a decrease in the superfluid density and therefore suppress
the superconductivity. The phase diagram of $T^*$, $T^{\nu}$, and
$T_{c0}$ versus Zn content is presented and discussed.
\end{abstract}

\maketitle

\section{\label{sec:level1}Introduction}

Effect of impurity doping is an important issue to discuss when
considering the mechanism of high $T_{c}$ superconductivity. In
YBa$_{2}$Cu$_{3}$O$_{7-\delta}$ (YBCO) system, divalent Zn ions
are believed to selectively substitute the planar Cu sites, which
causes a sharp drop of $T_c$. The effects of a nonmagnetic
impurity on the superconductivity have been theoretically
predicted for a d-wave superconductivity\cite{Maki95,Radtke93}.
Experimental studies, including NMR\cite{Alloul,Julien},
resistivity\cite{Fukuzumi}, surface impedance\cite{Panagopoulos},
electronic specific heat\cite{Loram}, optical
spectra\cite{Basov,Wang}, neutron scattering\cite{Kakurai}, muon
spin rotation\cite{Bernhard,Uemura}, and Raman-scattering
studies\cite{Altendorf,Matic,Limonov0,Limonov1}, have been
performed to investigate the effect of Zn doping in YBCO system.
These studies have suggested the pair-breaking effect and/or the
decrease in the superfluid density due to Zn doping accounts for
the radical suppression of superconductivity.

Recent Nernst effect measurements on high-\emph{T}$_{c}$
superconducting cuprates (HTS) have shown very surprising results
\cite{zax,ywang1,ywang2}. The Nernst effect is the appearance of a
transverse electric field $E_y$ in response to a temperature
gradient $-\nabla T\|\hat{x}$, in the presence of a perpendicular
magnetic $\mathbf{H}\|\hat{z}$ and under open circuit conditions.
The Nernst effect is usually small in the normal state of metals
where transport by quasi-particles is dominant. However, for a
type-II superconductor (in the vortex-liquid state), a new set of
excitations - vortices- are driven by temperature gradient
$-\nabla T\|\hat{x}$. Vortices diffuse down the gradient with
velocity $\mathbf{v}\|\hat{x}$. As each vortex core cross the line
between a pair of transverse voltage electrodes, the $2\pi$ phase
slip of the condensate phase leads to a Josephson $\mathbf{E}$
field given by $\mathbf{E}=\mathbf{B}\times\mathbf{v}$, which is
called vortex Nernst effect.

The experiments by Ong and collaborators \cite{zax,ywang1,ywang2}
have uncovered a large Nernst signal in the non-superconducting
state of hole-doped cuprates, at temperatures well above the
critical temperature $T_c$. The effect is particularly pronounced
in underdoped samples, extending well into the "pseudogap" region
of the cuprate phase diagram. The authors have interpreted this
anomalous Nernst signal above $T_c$ as evidence for vortex-like
excitations, and suggest that it is related to the pseudogap or
some interaction between the pseudogap state and the
superconducting state. This discovery has inspired a revisit to
the theory of superconducting fluctuations in the cuprates. The
conjecture that significant superconducting fluctuations in the
pseudogap region give rise to the large Nernst signal is accord
with the idea on the pseudogap proposed by Emery and Kivelson
\cite{Emery} that attributes its various anomalies to fluctuating
superconductivity. Kontani suggests that the behaviors of Nernst
signal below the pseudogap temperature ($T^*$) can be explained as
the reflection of the enhancement of the d-wave superconudcting
fluctuations and antiferromagnetic (AF)fluctuations, without
assuming thermally excited vortices\cite{Kontani}. Ussishkin
\emph{et al.}\cite{Ussishkin} calculate the contribution of
Gaussian superconducting fluctuations to the transverse
thermoelectric response above $T_c$ and find that the Gaussian
fluctuations are sufficient to explain the Nernst effect in the
optimally doped and overdoped samples. Moreover the importance of
the phase fluctuation of the complex order parameter $\psi$ has
been taken into account in theoretic models\cite{Emery, Timm}. In
the work of Carlson \emph{et al.}\cite{Carlson}, the fluctuations
in the phase of the order parameter would dominate the Nernst
signal up to a certain temperature above $T_{c}$.

The possibility that other exotic excitations in a strongly
correlated state cause the anomalous Nernst effect is not
excluded. For example, Weng and Muthukumar\cite{Weng} suggest that
in the description of spin-charge separation based on the phase
string theory of the t-J model, thermally excited spinons destroy
phase coherence, leading to a new phase characterized by the
presence of free spinon vortices at temperatures $T_c < T < T_v$.
The temperature scale $T_v$ at which holon condensation occurs
marks the onset of the pairing amplitude and is related to the
spin-pseudogap temperature $T^*$. The phase below $T_v$, called
the spontaneous vortex phase, shows novel transport properties
before phase coherence sets in at $T_c$, and the Nernst effect is
regarded as an intrinsic characterization of such a phase.

In this paper, we report the measurements of resistivity, Hall
effect and Nernst effect on the Zn doped
YBa$_{2}$Cu$_{3}$O$_{7-\delta}$ epitaxial thin films. We find that
Zn doping induces significant decrease of the vortex Nernst signal
in mixed state, the onset temperature (T$^{\nu}$) of anomalous
Nernst signal above $T_{c0}$, as well as the superconducting
transition temperature (\emph{T}$_{c0}$). Our results can be
understood in the scenario that Zn doping leads to a decrease in
superfluid density. The phase diagram of $T^*$, $T^{\nu}$, and
$T_{c0}$ versus Zn content is presented and discussed.

\section{\label{sec:level1}Experimental}

The c-axis-oriented epitaxial
YBa$_{2}$(Cu$_{1-x}$Zn$_{x}$)$_{3}$O$_{7-\delta}$
($x=0,0.005,0.01,0.02$) thin films were grown by pulsed laser
deposition (PLD) method on LaAlO$_{3}$ substrates which were cut
into a rectangle dimension of 10$\times$5 mm$^2$. The Zn content
was determined by the composition of the targets which were
prepared by standard solid state reaction. The temperature of the
substrates was typically $720\,^{\circ}\mathrm{C}$, and oxygen
pressure was 25 Pa during the deposition. The thickness of the
films was estimated to be about 200 nm according to the deposition
time. To get optimal oxygen content, the samples were annealed at
$500\,^{\circ}\mathrm{C}$ for half an hour under 1 atm pure oxygen
and $\delta$ was estimated to be less than 0.05. X-ray-diffraction
shows that the films are c-axis oriented perpendicular to the
substrate surface.

For electric resistivity and Hall effect measurements, six golden
electrodes were deposited on each film. The in-plane resistivity
$\rho(T)$ was measured by standard four-probe method. The Hall
coefficient $R_{H}(T)$ was measured under a magnetic field of 5T
parallel to the c axis of the film. We define the (x, y) plane as
the conducting ab plane of the film samples. In the set-up of
Nernst effect measurement, a temperature gradient of about 3K/cm
is applied along longitudinal direction (x-direction), and the
magnetic field $H$ is applied along z-direction (perpendicular to
the thin film surface). Thus the Nernst electric field is along
y-direction, which was measured by Keithley Model 2182
Nano-voltmeter. The temperature gradient was measured by two small
Cernox bare-chip thermometers (CX-1050-BR) which were attached to
the two ends of the sample. A small heater is on the free end of
the sample. All the measurements are based on a Quantum Design
PPMS-9 system with the temperature drift less than 0.05\%. The
Nernst coefficient $\nu$ is defined as

\begin{eqnarray}
\nu\equiv{E_{y}\over -\nabla_{x}\emph{T}\cdot\emph{B}_{z}}
\label{eq:one}.
\end{eqnarray}
We also define the Nernst signal as
$e_{y}$$\equiv$$E_{y}$/$(-\nabla_{x}\emph{T})$. The Nernst signal
was measured at positive and negative field polarities, and the
difference of the two polarities was taken to remove any
thermopower contribution. Since $e_y$ is not linear with $B$ as $T
< T_{c0}$, the Nernst coefficient $\nu$ in the mixed state was
calculated from the initial slope of $e_y$ versus $B_z$.

\section{\label{sec:level1}Results and Discussion}

\begin{figure}
\includegraphics[width=8cm]{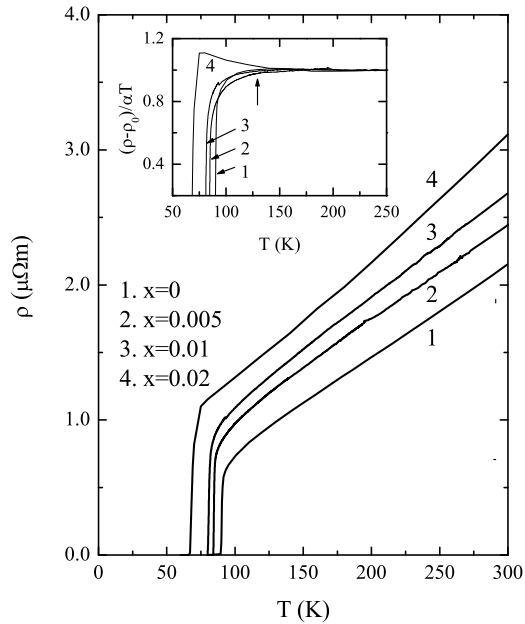}
\caption{\label{fig1} the temperature dependence of in-plane
resistivity for Zn doped
YBa$_{2}$(Cu$_{1-x}$Zn$_{x}$)$_{3}$O$_{7-\delta}$
($x=0,0.005,0.01,0.02$) thin films. The inset of Fig.~\ref{fig1}
shows the plot of $(\rho-\rho(0))$/$\alpha$$T$, versus
temperature, where $\rho(0)$ is the $T$=0 intercept of the
extrapolated $T$-linear high temperature curve and $\alpha$ the
slope of the linear part of the resistivity.}
\end{figure}

Fig.~\ref{fig1} shows the temperature dependence of in-plane
resistivity, $\rho$, for Zn doped
YBa$_{2}$(Cu$_{1-x}$Zn$_{x}$)$_{3}$O$_{7-\delta}$
($x=0,0.005,0.01,0.02$) thin films. The resistivity shows linear
temperature dependence at high temperature for all 4 samples, and
it deviates downward for $x \leq 0.01$ when $T$ approaches
$T_{c}$. For the sample with $x$=0.02, the resistivity shows a
small upturn at low temperature before it drops to zero at
$T_{c0}$. Such a low-temperature upturn in $\rho$ is usually
ascribed to the localization effects. The zero-point
superconducting transition temperature, $T_{c0}$, is 90, 84, 79,
and 67 K for $x$=0, 0.005, 0.01, and 0.02 determined from the
resistive transitions, consistent with the results in literatures.
The crossover temperature $T^{*}$ at which $\rho (T)$ deviates
downwards from linear behavior corresponds to the onset of the
pseudogap opening \cite{Ito}. To show $T^{*}$ clearly, the inset
of Fig.~\ref{fig1} shows the plot of $(\rho-\rho(0))$/$\alpha$$T$,
versus temperature, where $\rho(0)$ is the $T$=0 intercept of the
extrapolated $T$-linear high temperature curve and $\alpha$ the
slope of the linear part of the resistivity. Due to the
localization effect for $x$=0.02, $T^{*}$ can not be determined
reliably from the resistivity measurement. Contrast to underdoped
YBCO whose $T^*$ increases as $T_{c}$ decreases with oxygen
depletion, $T^*$ is nearly unaffected by Zn doping although
$T_{c0}$ drops drastically. The insensitivity of $T^{*}$ to Zn
impurities has been well documented in the literatures
\cite{Walker,Abe} for both fully oxygenated and oxygen depleted
YBCO. However, it should be noted that the deviation of $\rho (T)$
below $T^*$ becomes smaller as $x$ increases, which means that the
pseudogap is filled up by Zn doping.

\begin{figure}
\includegraphics[width=8cm]{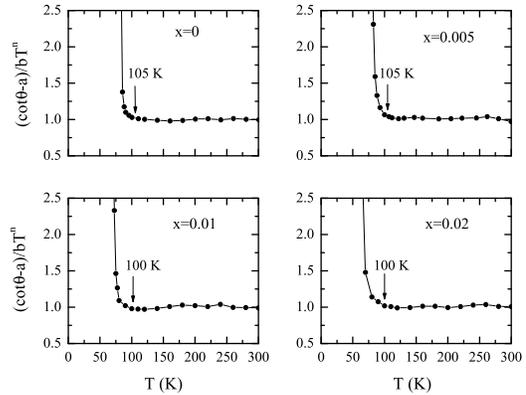}
\caption{\label{fig2} The plots of (cot$\theta$-$a$)/$bT^n$ versus
$T$, to show the deviation from the high-temperature behavior
$cot\theta$ = $a$+$b$$T^n$. The deviation at $T_1$ is marked by
arrows.}
\end{figure}

The temperature dependence of Hall coefficient, $R_{H}$, was also
measured for Zn doped
YBa$_{2}$(Cu$_{1-x}$Zn$_{x}$)$_{3}$O$_{7-\delta}$
($x=0,0.005,0.01,0.02$) thin films. The Hall coefficient in normal
state increases slightly and the temperature dependence of $R_{H}$
becomes a little weaker as Zn content increases. According to the
previous studies \cite{Xiao,Cooper,Chien,Walker}, although Zn
doping leads to an enhancement of the in-plane scattering rate,
but the hole carrier concentration and the profile of $R_{H}(T)$
curves change little if the Zn content is low enough. Generally
the Hall angle in the normal state satisfies the relation
cot$\theta$ =$a+bT^2$ which can be interpreted in the two
relaxation times picture\cite{Chien,Anderson}. The peak in $R_{H}$
around 100 K has been attributed to the change in the $T^2$
dependence of cot$\theta$. Many studies have suggested that the
peak in $R_{H}$ or the change of cot($\theta$)($T$) is also
associated with the opening of the pseudogap
\cite{Ito,Bucher,Hwang,Ott,Matthey}, but $T_1$ usually falls
between $T_c$ and $T^*$ (determined from $\rho(T)$). For our Zn
doped YBCO samples, cot$\theta$ in normal state can be fitted by
the function $a+bT^n$ with $n$ close to 2. In Fig.~\ref{fig2} the
temperature dependence of (cot$\theta$-$a$)/$bT^n$ is shown for
the four samples. The fitting parameters are $a$ = -61.56, 86.27,
217.65, 292.90, $b$ = 0.1055, 0.0769, 0.146, 0.125, and $n$ =
1.85, 1.89, 1.75, 1.76, for $x$=0, 0.005, 0.01, 0.02,
respectively. Way above $T_c$, (cot$\theta$-$a$)/$bT^n$ is quite a
constant of 1 over a wide temperature range. The temperature at
which (cot$\theta$-$a$)/$bT^n$ deviates from 1 is shown by an
arrow for each Zn-doping case. We define this temperature scale as
$T_1$. Although $T_1$ ($\sim$ 100 K) is lower than $T^*$, both are
almost independent on the Zn content. However, since $T_1$ is
always much lower than $T^*$, it is suggested that $T_1$ defines a
new crossover temperature which is not related to
pseudogap\cite{Matthey}, or there are two temperature scales for
the pseudogap\cite{Abe}. Matthey \emph{et al.}\cite{Matthey}
proposed that this temperature scale, $T_1$, may be related to the
superconducting fluctuation or vortex-like excitation above $T_c$
detected by Nernst effect measurement, which will be shown and
discussed below.

\begin{figure}
\includegraphics[width=8cm]{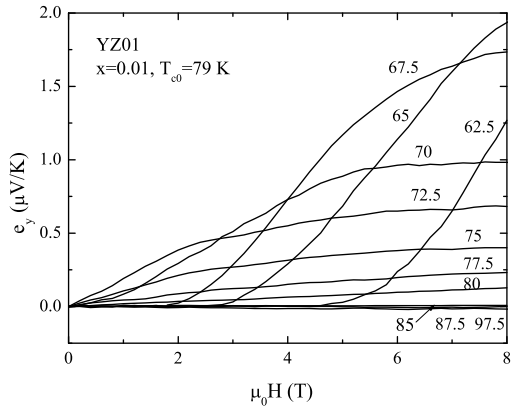}
\caption{\label{fig3} The field dependence of the Nernst signal in
Zn doped YBCO (sample YZ01, $x$=0.01) at fixed $T$ from 62.5 to
97.5 K. For $T$ just below $T_{c0}$ (=79 K), $H_m$ $\simeq$ 0, and
$e_y$ vs $H$ shows a very pronounced negative curvature. $H_m$
increases as $T$ decreases.}
\end{figure}

\begin{figure}
\includegraphics[width=8cm]{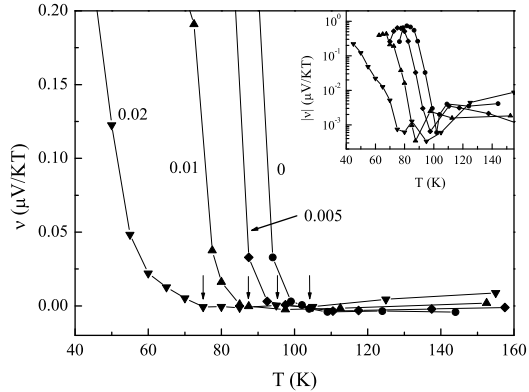}
\caption{\label{fig4} The temperature dependence of $\nu$ for Zn
doped YBa$_{2}$(Cu$_{1-x}$Zn$_{x}$)$_{3}$O$_{7-\delta}$
($x=0,0.005,0.01,0.02$) thin films. The temperature at which the
Nernst coefficient deviates from normal state background is shown
by an arrow for each case. The inset shows the semi-logarithmic
plot of $|\nu|$ vs $T$.}
\end{figure}

\begin{figure}
\includegraphics[width=8cm]{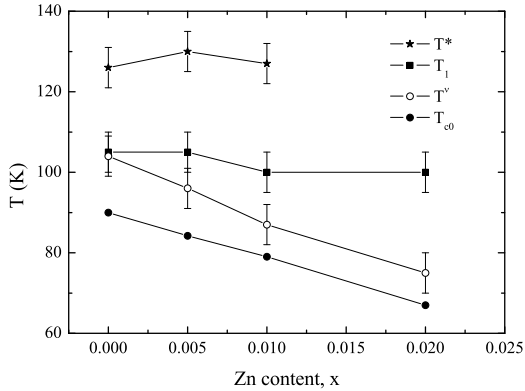}
\caption{\label{fig5} The variation of $T^*$, $T_1$, $T^{\nu}$,
and $T_{c0}$ with Zn content, $x$.}
\end{figure}

The Nernst effect of all the Zn doped YBCO was measured. Figure
~\ref{fig3} shows the traces of the Nernst signal $e_{y}$ versus
the applied magnetic field $H$ of a typical sample with $x$=0.01
($T_{c0}$=79 K) taken at fixed temperatures. At temperatures well
below $T_{c0}$, all the curves have the characteristic features of
the vortex Nernst effect: the signal $e_{y}$ remains zero in the
vortex lattice state where all the vortices are pinned; after a
first order vortex solid to liquid phase transition at $H_m$, the
motion of a large number of vortices leads to a sharp increase of
$e_{y}$; it tends to reach a maximum at a higher characteristic
field scale $H^*$ ($H^*$ is beyond the maximum of the applied
magnetic filed in some low temperatures). With increasing
temperatures, $H_m$ tends to smaller values and disappears as $T$
is close to $T_{c0}$. As $T$ is above $T_{c0}$, the Nernst signal
is still a sizeable fraction of low-$T$ values. At higher $T$, the
signal decreases gradually, approaching a straight line of
negative slope, which we identify with the background signal
$\nu_n$ from the normal charge carriers (holes). All the samples
studied exhibit similar traces of $e_y$ versus $H$. The
temperature dependence of Nernst coefficient $\nu$ is shown for
the samples with $x$ =0, 0.005, 0.01, and 0.02 in Figure
~\ref{fig4}. The inset of Figure ~\ref{fig4} shows the absolute
value of $\nu$ versus $T$ in semi-logarithmic scale. The arrows
indicate the onset temperature $T^{\nu}$ at which the anomalous
Nernst signal is resolved from the high-temperature normal state
background $\nu_n$. The Nernst signal in normal state is small and
almost temperature independent, similar to the underdoped HTS. The
contribution of normal charge carriers to the Nernst signal is
usually very small in cuprates due to so-called "Sondheimer
cancellation"\cite{Sondheimer}. It should be mentioned that the
contribution of the Cu-O chains to Nernst signal can not be
isolated in our measurement because of the twinned samples.
However, we think that the intrinsic Nernst coefficient of ab
plane should still be obtained because the Nernst coefficient in
normal state is mainly related to the off-diagonal components of
electric conductivity tensor and Peltier conductivity tensor and
Ando's group has recently confirmed that the Onsager's reciprocal
relation of resistivity [$\rho_{yx}(H)$=$\rho_{xy}(-H)$] still
holds for YBCO regardless of the conduction of the Cu-O chains
\cite{Ando}. With decreasing temperatures, $\nu$ deviates from the
high-temperature background and increases quickly at the onset
temperature, $T^{\nu}$, then reaches a maximum below $T_c$, and
finally decreases due to flux-pinning. In the temperature range
between $T_{c0}$ and $T^{\nu}$ there exists anomalously large
Nernst signal, which has been interpreted as evidence for
vortex-like excitations or strong superconducting fluctuations in
this region. Usually $T^{\nu}$ is lower than $T^*$ in underdoped
HTS, but it scales with $T^*$. However, in Zn doped YBCO,
$T^{\nu}$ decreases quickly with $x$ as $T_{c0}$ does. We show the
variation of $T^*$, $T_1$, $T^{\nu}$, and $T_{c0}$ with Zn
content, $x$, in Figure ~\ref{fig5}. It can be seen that while
$T_1$ and $T^*$ change little, $T^{\nu}$ decreases sharply as
$T_{c0}$ does, and the interval between $T_c$ and $T^{\nu}$
remains almost unchanged as $x$ increases.

It should be noted that the Nernst signal below $T^\nu$ decreases
as Zn content increases. The  maximum of $\nu$ decreases
monotonously with $x$ while the high-temperature background of
$\nu$ does not change significantly. This result is content with
the conclusion that there is a drastic effect of Zn doping on the
superfluid density $n_s$ probed by $\mu$SR measurements
\cite{Uemura}. NMR\cite{Alloul,Julien} and neutron
study\cite{Kakurai} have found that Zn doping induces a localized
magnetic moment in the CuO$_2$ plane although Zn is itself a
nonmagnetic impurity. Furthermore, it has been found that the
superfluid density detected by $\mu$SR probe decreases with Zn
content and a "Swiss cheese" model in which charge carriers within
an area $\pi\xi^2_{ab}$ around each Zn impurity are excluded from
the superfluid has been proposed\cite{Uemura}. Recent results of a
scanning tunneling microscope (STM) measurement\cite{STM} have
also revealed that superconductivity is locally destroyed around
Zn sites. Such an spatial variation of order parameter is not
expected in the conventional homogeneous picture. Roughly
speaking, the Nernst signal $e_y$ is determined by the product of
vortex density $n_{\phi}$ and moving velocity $v_{\phi}$ of
vortex. The decrease of superfluid density leads to a decrease of
$n_{\phi}$, and the strong suppression of order parameter within a
given area around Zn impurities obstructs the moving vortices like
disorder potential. Therefore the sharp decrease in the maximum of
Nernst signal with Zn content is consistent with the "Swiss
cheese" model.

In underdoped La$_{2-x}$Sr$_x$CuO$_4$, $T^{\nu}$ is about half of
$T^*$ and the interval between $T^{\nu}$ and $T_{co}$ is as large
as 100 K for $x$=0.10\cite{zax}. Such a large regime of
superconducting fluctuations can not be understood in conventional
superconducting fluctuation theory. This result has been regarded
as the evidence supporting the precursor superconductivity
scenario to explain the pseudogap
phenomena\cite{zax,ywang1,ywang2}. However, in Zn doped YBCO, the
four temperature scales, $T^*$, $T_1$, $T^{\nu}$, and $T_{c0}$,
can be divided into two categories according to their dependence
on Zn content. $T^*$ and $T_1$ are in one category: both are
almost independent on $x$. In the other hand, $T^{\nu}$ and
$T_{c0}$ are in the other category: both decrease quickly with
$x$. $T^{\nu}$ and $T^*$ seems to be unrelated, which is opposite
to the precursor superconductivity scenario. Namely, the anomalous
Nernst signal above $T_{c0}$ may not be related to the pseudogap.
We suggest two possibilities to explain the contradictory behavior
of $T^{\nu}$ and $T^*$. One possibility is that there do not exist
preformed Cooper pairs in pseudogap region, and $T^*$ is the onset
temperature of the fluctuations of some other types, such as
antiferromagnetic fluctuations, charge density waves, or
electronic phase separation (e.g., the stripe scenario), which
compete or coexist with superconductivity\cite{Tallon1, Tallon2}.
In this case Zn impurities only suppress the superconductivity and
leave the fluctuation of other type unchanged. $T^{\nu}$ which is
just the onset of the superconducting fluctuation decreases with
Zn content as $T_{c0}$ does. However, the scale of $T^{\nu}$ with
$T^*$ and the large interval between $T^{\nu}$ and $T_{co}$ in
underdoped cuprates are difficult to be understood in this
picture. The other possibility is that $T^*$ is the onset of
precursor superconductivity which does not destroyed by Zn
impurities, while the sharp decrease in the superfluid density and
the spatial variation of order parameter due to Zn doping makes
the vortex-like excitations above $T_{c0}$ weak and undetectable
by Nernst effect measurement, leading to a decrease of $T^{\nu}$.
The fact that the sharp decrease in the maximum of Nernst signal
with Zn content below $T_{c0}$ also supports this picture. We
suggest that Zn doping does not have much influence on the
pseudogap opening, and therefore neither $T_1$ nor $T^*$ changes
much with Zn content. However, Zn impurities cause a notable
decease in the superfluid density, and therefore suppress both
$T^{\nu}$ and $T_{c0}$.

\section{\label{sec:level1}Conclusion}

In conclusion, we have studied the transport properties, including
resistivity, Hall effect and Nernst effect in a series of Zn doped
YBa$_{2}$(Cu$_{1-x}$Zn$_{x}$)$_{3}$O$_{7-\delta}$ epitaxial thin
films. It is found that the pseudogap temperature $T^*$ determined
from the temperature dependence of resistivity and the temperature
scale $T_1$ determined from the $T^n$ dependence of cot$\theta$
remain unchanged, meanwhile the onset temperature $T^{\nu}$ of
vortex-like excitation above $T_{c0}$ determined from Nernst
effect drops sharply with increasing Zn content as the
superconducting critical temperature $T_{c0}$ does. We also find
that the vortex Nernst signal in mixed state decreases quickly
with increasing $x$. The variations of $T^*$ and $T^{\nu}$ with
$x$ might be understood in the picture that Zn doping does not
destroy the precursor superconductivity, but causes a sharp
decrease in the superfluid density and spatial variation of the
order parameter, as suggested in the "Swiss cheese" model.

\begin{acknowledgments}
The authors would like to thank C. M. Feng for help on transport
measurements and thank G. H. Cao for fruitful discussions. This
work was supported by the National Natural Science Foundation of
China (Grant No. 10225417) and the Ministry of Science and
Technology of China (project: NKBRSF-G1999064602).
\end{acknowledgments}

\end{document}